\documentclass[preprints,article,accept,moreauthors,pdftex]{Definitions/mdpi}
\firstpage{1} 
\pubvolume{1}
\issuenum{1}
\articlenumber{0}
\pubyear{2022}
\copyrightyear{2022}
\datereceived{} 
\dateaccepted{} 
\datepublished{} 
\hreflink{https://doi.org/} 
\Title{Gas bubble photonics: Manipulating sonoluminescence
    light with fluorescent and plasmonic nanoparticles}

\TitleCitation{Gas bubble photonics}

\usepackage{textcomp}
\usepackage{bm}
\usepackage{comment}

\Author{Ivan S.~Maksymov \orcidA{}}

\AuthorNames{Ivan S.~Maksymov}

\AuthorCitation{Maksymov, I.~S.} 

\address{%
\quad Optical Sciences Centre, Swinburne University of 
Technology, Hawthorn, Victoria 3122, Australia\\} 

\corres{Correspondence: imaksymov@swin.edu.au ; Tel.: +61-3-3921-4805}

\abstract{Oscillations of gas bubbles in liquids irradiated with acoustic pressure
waves may result in an intriguing physical phenomenon called sonoluminescence,
where a collapsing bubble emits the light in a broad optical spectral range. However,
the intensity of the so-generated light is typically weak for practical purposes.
Recently, it has been demonstrated that nanoparticles can be used to increase the
efficiency of sonoluminescence, thereby enabling one to generate the light that
is intense enough for a number of applications in photonics, biomedicine and material
science. In this article, we review the latest achievements in the field of
nanoparticle-enhanced sonoluminescence and showcase the perspectives of their
practical applications.}


\begin{document}
\section{Introduction\label{sec:1}}

A soap bubble floating in the air, air bubbles rising from a scuba diver in water 
and bubbles of gas rising in a soft drink are just a few examples of one remarkable
physical phenomenon that most of us know since our childhood \cite{Bre95, Lau10}.
We can see an air bubble in water because the optical refractive index of air is
approximately 1 but that of water is approximately 1.33, which, according to
Snell's law \cite{Bor97}, allows identifying a bubble due to the effect of
optical refraction \cite{Rac22} and internal reflection processes \cite{Mil20}.

\begin{figure}
  \includegraphics[width=.7\textwidth]{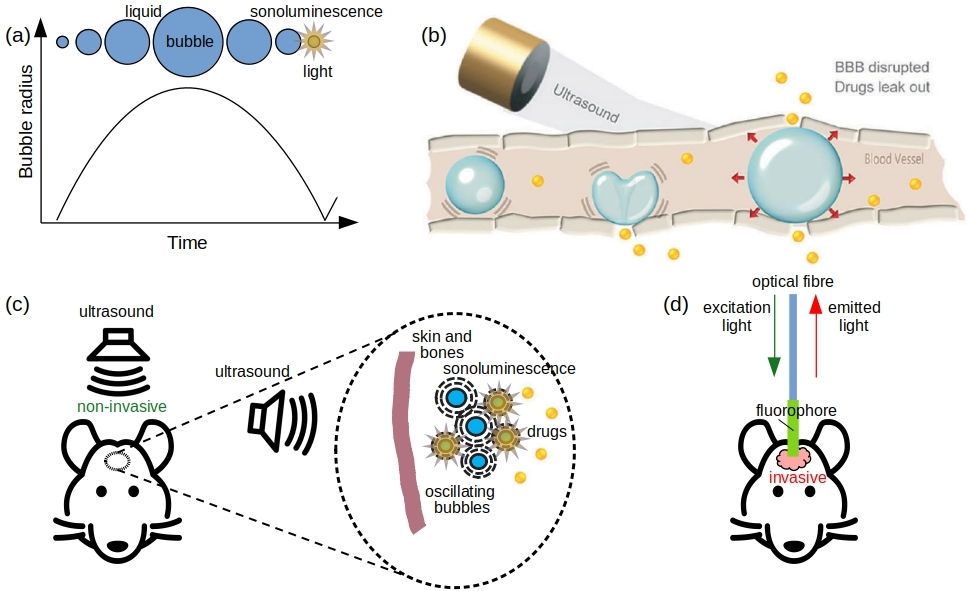}
  \caption{
    (\textbf{a})~Schematic radius-vs-time diagram for a gas bubble
    oscillating in the bulk of water. Bubble shapes at different times
    are shown above the curve during a single oscillation cycle. The
    pressure inside a bubble is high at the beginning and at the end
    of the oscillation cycle and is low in the middle.
    (\textbf{b})~Illustration of the principle of drug delivery
    through the blood-brain barrier (BBB) using gas bubbles inside a
    small blood vessel of the brain. Bubble oscillations are driven by
    an ultrasound wave transducer located outside the head and the
    drug (small spheres in the picture) pass through the wall of the
    blood vessel when the bubble either reaches its maximum radius or
    collapses and forms a water jet. Reproduced from \cite{Kang} under
    the terms of a Creative Commons license.
    (\textbf{c})~Sketch of a potential scheme of drug delivery with gas
    bubbles and subsequent drug photoactivation with sonoluminescence UV
    spectral range light. Note a non-invasive nature of this procedure 
    since ultrasound waves needed to trigger the bubble collapse and
    sonoluminescence can safely pass through the skin, bones and bodily
    fluids.
    (\textbf{d})~Illustration of an idealised fluorescence-based fibre-optic 
    probe that can be used for imaging, sensing and drug photoactivation
    inside a living body. Note that using this technique requires inserting 
    an optical fibre into the body, i.e.~this approach requires a surgical
    intervention.   
    \label{Fig_intro}}
\end{figure}

However, in many natural phenomena and human-made technologies bubbles are rather
undetectable with a human eye. Indeed, although the ability of bubbles to affect
light has been exploited in atmospheric and marine sciences \cite{Shifrin, Han85}
and optical sensing systems \cite{Xu18, Mak19}, bubbles have mostly been the subject
of extensive studies in the fields of acoustics, fluid dynamics, biomedicine,
chemistry and material science, where their size falls mostly within the micron 
range \cite{Tsu14}. For example, artificial microbubbles have been employed in
modern biomedical technologies including drug delivery \cite{Bur15, Kang, Den18} and
contrast-enhanced medical ultrasound imaging procedures \cite{Lin04, Pos11, Kang}.
Ultrasound contrast agents rely on the ability of spherical bubbles to
periodically expand and collapse (oscillate) [Fig.~\ref{Fig_intro}(a)] when they
are irradiated with ultrasound waves. Commercially available contrast media are
specially-prepared lipid-coated microbubbles that can be administered intravenously
to the systemic circulation \cite{Lin04, Bur15, Kang, Den18} and that change the
way ultrasound waves are reflected from bodily tissues and fluids. Oscillating
bubbles of similar size and composition can be used to deliver drugs to hard-to-reach
places inside a living body [Fig.~\ref{Fig_intro}(b)], which is, for example, the
case of drug delivery through the blood-brain barrier (BBB) that protects the brain
against circulating toxins and pathogens causing its infections \cite{Dan15}. We
note that the bubble in the middle of the blood vessel in Fig.~\ref{Fig_intro}(b)
highlights another physical regime of bubble oscillation and collapse suitable for
opening BBB and delivering drugs through it, where the bubble collapses non-spherically
forming a powerful jet \cite{Blake_book, Ohl15, Man16}. The physics of non-spherically
collapsing bubbles is more complex but, in general, similar to that of spherically
collapsing bubbles that we discuss in continuation. 

Apart from their applications in biomedicine, bubbles have also been used for texture
tailoring in food industry \cite{Dut04}, natural gas recovery in petroleum industry
\cite{May18}, material synthesis in material sciences \cite{Kha12}, lab-on-a-chip
devices \cite{Has12}, wastewater treatment systems \cite{Gog08}, sonochemistry
(enhancement and alternation of chemical reactions by means of ultrasound)
\cite{Mas03, Tan11}, sonoprocessing \cite{Mas03, Rub16} and underwater acoustic
communication \cite{Pre03, Mak22}. Moreover, it has been hypothesised that acoustic
and fluid-mechanical properties of bubbles in the primordial ocean might contribute
to the origin of life on Earth \cite{Kal17} and that the presence of bubbles in the
brain and their collapse might be associated with blast-induced neurotraumas
\cite{Cer17, Bar20}. The physics of gas bubbles is also relevant to processes
taking place during underwater explosions \cite{Col48}.

However, the aforementioned list of properties and practical applications of
bubbles is complete. In this article we focus on the well-established
capability of an oscillating bubble to emit light in a broad optical spectral
range via the process called sonoluminescence \cite{Put00, Bre02, Far11, Bor15,
  Yas21}. As schematically shown in Fig.~\ref{Fig_intro}(a), while a bubble in
the bulk of liquid periodically expands and collapses under the pressure of an
ultrasound wave, at certain experimental conditions the shrinkage of the bubble
can result in the emission of pulses of light \cite{Put00, Bre02}. The intensity
of the emitted light is typically higher in experiments involving individual
bubbles \cite{Bre02} compared with the case of multibubble sonoluminescence
\cite{Yas21}, where the brightness of sonoluminescence is adversely affected by
interactions between neighbouring bubbles. In fact, while the light emitted by a
single air bubble oscillating in a water-filled glass flask can be seen with the
naked eye in a darkened room when using low-power kHz-range ultrasound pressure
waves \cite{Bre02}, visible multibubble sonoluminescence is possible mostly
using high-frequency high-intensity ultrasound waves and bubbles that contain
noble gases such as xenon \cite{Put00}.      

Moreover, since sonoluminescence light has a significant spectral ultraviolet
(UV) component \cite{Put00, Bre02, Boy20} but bubble oscillations resulting
in sonoluminescence can be triggered by ultrasound waves that can pass through
the body without noticeable attenuation, it has been argued \cite{Beg19, Can20}
[Fig.~\ref{Fig_intro}(c)] that sonoluminescence light can be exploited to
noninvasively photoactivate drugs inside a living body in hard-to-reach
organs such as the brain. Such ideas are of immediate relevance to the current
research efforts in the field of photopharmacology that aims to develop smart
drugs that, through the incorporation of molecular photoswitches, enable the
remote spatial and temporal control of bioactivity by light \cite{Vel14, Weg17}.
This concept is particularly beneficial in the treatment of bacterial infections,
where there exists the need of controlling the side effects of antibiotics. However,
its practical implementation faces several problems since modern light-responsive
drugs mostly rely on UV light, which is a problem because UV radiation is
strongly attenuated by bodily fluids and tissues but increasing its intensity
results in adverse phototoxic effects on biological cells. These challenges
motivate the research on novel drugs that could be activated using red light
that falls within the so-called therapeutic window \cite{Weg17}, where light
has its maximum depth of penetration in tissue \cite{Smi09}. However, using 
sonoluminescence one might generate UV light directly inside the body, thus 
resolving the problem of light delivery to the body interior from an
external source of UV radiation and also enabling one to use the already  
discovered UV-sensitive drug compositions. Yet, it has been theoretically
shown that sonoluminescence UV light generated inside an internal body organ
may have an additional antibacterial effect \cite{Boy20} that can improve the
overall performance of UV-activated drugs. 

More broadly speaking, using sonoluminescence in the field of biomedicine
may be advantageous compared with well-established optical sensing, imaging
and drug activation techniques \cite{review_CNBP}, including fluorescence-based
methods \cite{Ton12, Reineck}, where the use of laser beams and optical
fibre probes [Fig.~\ref{Fig_intro}(d)] implies that a surgical intervention
is needed to access an internal organ such as the brain. On the contrary, a 
sonoluminescence-based approach may be fully noninvasive since it could 
rely on advances in the well-established field of conventional medical and
transcranial ultrasound imaging methods, where imaging of internal organs
and noninvasive functional brain studies are enabled by microbubble-based
contrast agents \cite{Err16}.

Fluorescence-based techniques rely on fluorophores---molecules or
nanoparticles---that can be applied directly to live specimens to investigate
the interior of a cell {\it in situ} or an internal body organ {\it in vivo}
\cite{review_CNBP, Reineck}. The light absorption by a fluorophore is followed
by re-emission of a photon with a lower energy and a longer wavelength. Hence,
fluorescence-based techniques separate the re-emitted lower intensity light
from its excitation counterpart creating an image or enabling one to detect
chemical composition \cite{Pur15} and temperature \cite{Mus16, review_CNBP}.

However, despite their technological maturity, fluorescence-based approaches
have several other notable disadvantages apart from the aforementioned invasive
nature of fluorescent probes. Firstly, the operating time of fluorescent molecules,
such as those that cover the tip of the probe in Fig.~\ref{Fig_intro}(d) \cite{Pur15},
is often insufficient for real-time monitoring due to the physico-chemical effect
called photobleaching \cite{Reineck}. This problem cannot be simply resolved by
increasing the intensity of the excitation light since doing so may cause photodamage
to biological cell \cite{Zab20}. While novel fluorescent nanoparticles that
require less intense excitation light and that are less prone to photobleaching
have been developed \cite{Reineck}, their long-term effect on the cell physiology
has not been studied in much detail yet. Therefore, their practical use remains
very limited \cite{Reineck}. Yet, the use of fluorescence-based approaches during
{\it in vitro} fertilisation (IVF) poses significant problems---since their effect
on the embryo development is unknown, a direct contact of fluorophores with an
embryo is ethically unsound and is not permitted in many regulatory
jurisdictions \cite{Pur15}. 

These disadvantages are not expected to exist in sonoluminescence-based
approaches. For example, continuing the discussion of the very important and
ethically challenging field of IVF, it is well-known that both low-intensity
ultrasound waves \cite{Whi15} and microbubble ultrasound contrast agents are
safe for obstetric imaging applications \cite{Yus22}. Therefore, it is plausible
that bubble-based techniques that produce light would be also suitable for
applications in this field. Similarly, sonoluminescence could be achieved
using bubbles that serve as ultrasound contrast agent in intravascular imaging
procedures \cite{Wan10}, where it has been demonstrated that the same bubbles
that provide contrast for ultrasound waves used for imaging purposed can be
laden with drugs \cite{Wil12_1, Dov13}. Subsequently, it is conceivable that 
sonoluminescence light could be used to photoactivate drugs that cover the 
surface of sonoluminescing microbubbles.          

\begin{figure}
  \includegraphics[width=.6\textwidth]{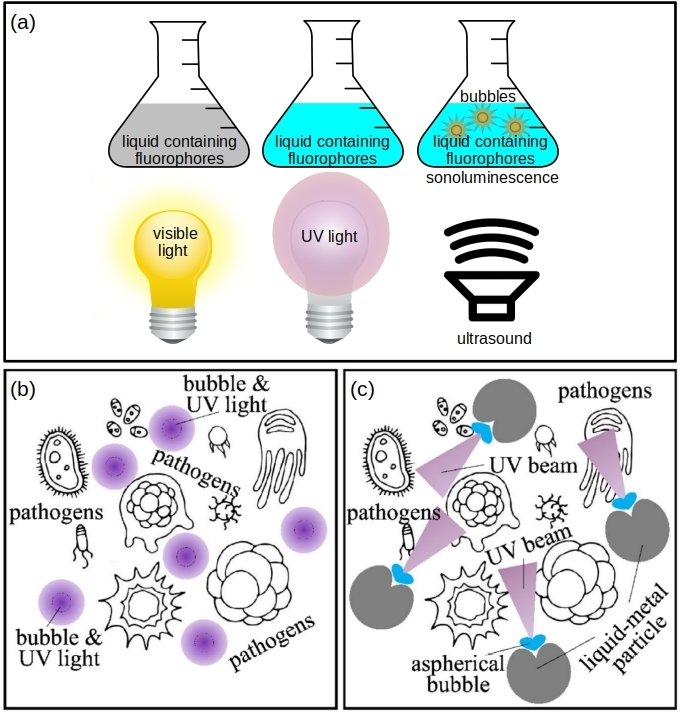}
  \caption{
    (\textbf{a})~Sketch of a potential proof-of-principle experiment involving 
    a flask filled with water that contains a fluorescent material. When the 
    flask is illuminated with visible light the liquid does not change its 
    colour. However, when the flask is illuminated with UV light the liquid
    changes its colour due to fluorescence processes. In the rightmost image,
    the liquid in the flask contains bubbles and it is irradiated with ultrasound,
    which leads to the emission of sonoluminescence light. Since the so-generated
    light has a significant UV component, fluorescence processes result in a 
    change in the colour of the liquid. Quinine---a flavour component of tonic
    water---may be used to realise this experiment since it is known to absorb
    UV light with the wavelength of around 350\,nm and to emit a bright blue light 
    at around 460\,nm. As shown in the main text, other suitable chemicals,
    including luminol, can also be employed.
    (\textbf{b,~c})~The intensity of sonoluminescence light can be enhanced using 
    plasmonic properties of metal nanoparticles that originate from coupling of 
    light to a resonant oscillation of conduction electrons at the metal surface
    \cite{Raether, Mak16}. This approach was theoretically demonstrated in 
    \cite{Boy20} in the context of disinfection of water contaminated by
    pathogens. It was shown that while the intensity of UV light emitted 
    by spherically collapsing gas bubbles is insufficient for UV germicidal 
    irradiation [panel~(b)], sonoluminescence associated with aspherical bubble
    collapse near a metal particle should result in the formation of a UV light
    beam [panel~(c)] that is intense enough for inactivating 99.9\% of most common
    pathogens. Reproduced from \cite{Boy20} under the terms of a Creative Commons
    license.\label{Fig2}}
\end{figure}

However, the implementations of the idea of sonoluminescence-activated drugs 
and of relevant proposals have thus far faced a number of fundamental physical 
and technical limitations---the intensity of the brightest sonoluminescence light
attainable in carefully planned and executed experiments is rather low compared
with that used in fluorescence-based and similar techniques. Indeed, in a typical
single-bubble sonoluminescence experiment involving a sub-millimetre-sized bubble
in a water-filled flask that serves as a resonator for kHz-range ultrasound waves,
the liquid has to be purified and degassed, the peak pressure amplitude of ultrasound
has to reach an optimal value but the bubble has to be created and placed in an exact
location inside the flask \cite{Put00, Bre02}. Given this complexity, for a long time
it was assumed that such experimental conditions could not be fulfilled in any
scenario of a bubble trapped in a biological fluid, where the chemical composition
of the fluid can be highly variable and there would be more many oscillating bubbles
that interact one with another \cite{Doinikov_book}. Nevertheless, recently it has
been shown that sonoluminescence can be observed under medically relevant conditions
using microbubbles commonly employed as contrast agents for ultrasound imaging
\cite{Beg19}. At the same time, recent works \cite{Boy20, Vig20, Son21} also
proposed several novel approaches to enhancement of sonoluminescence light intensity
to a level that is detectable by standard equipment and is also suitable for
photoactivation of drugs \cite{Vel14, Weg17}. In particular, it was shown that
the presence of nanoparticles, which does not affect acoustic properties of a bubble,
amplifies sonoluminescence by fluorescence-like \cite{Bro12, Vig20, Son21}
[Fig.~\ref{Fig2}(a)] and plasmonic \cite{Boy20} effects [Fig.~\ref{Fig2}(b,~c)].
Thus, these results strongly speak in favour of the concept sonoluminescence therapy
and photoactivation of drugs \cite{Can20}, especially because plasmonic nanoparticles
have played an important role in biomedical technologies \cite{Mak16} and, therefore,
their application in sonoluminescence-based technique should help resolve several
important problems such as biocompatibility and ethics of use in {\it in vivo} settings.

\section{Discussion\label{sec:2}}
In this section, we review the results of key works relevant to the mainstream 
discussion in this article. We note that multibubble sonoluminescence was
discovered in the 1930s \cite{Mar33, Fre34} and that this phenomenon has been studied 
in great detail since then \cite{Sus08, Yas21}. Single-bubble sonoluminescence was
discovered in the late 1980s \cite{Gai90, Bre02}, although according to some
sources \cite{Yas21} the very first observation of this effect was reported in
1962~\cite{Yos62}. One remarkable observation made in single-bubble sonoluminescence
measurements was the fact that the brightness of light emitted by a bubble was
detectable with the naked eye \cite{Bre02}. Yet, the equipment required to observe
this effect employed relatively simple electronics. Compared with a much low
brightness of light emitted in typical multibubble sonoluminescence experiments
that require more complex high-power acoustic power setups, the single-bubble
sonoluminescence effect immediately attracted considerable attention of researchers
due to its promising potential practical applications. Thus, it is plausible that
ideas of using sonoluminescence light for biomedical purposes could be expressed
at the same time. Furthermore, in the late 1980s several workers in the field of
bubble acoustics and fluid dynamics initiated the studied of cavitation processes 
using therapeutic ultrasound \cite{Pic88, Ume90} (also note a relevant earlier work,
where cavitation and sonochemiluminescence was investigated using high-kHz acoustic
pressure waves \cite{Sri61}). This fact also speaks in favour of the assumption
that the researchers aimed to use sonoluminescence for medical purposes in the 
1980s. On the other hand, according to \cite{Can20} the first evidence of a
biomedical application of sonoluminescence was given in the work \cite{Zha94}
that was published in a medical journal, where the authors proposed using
sonoluminescence in blood plasma to detect tuberculosis and cancer of the lungs.

However, even though the pioneering experiments on sonoluminescence employed
sophisticated experimental equipment, the progress in the field of scientific
instrumentation and data processing made in the past decades has enabled new
approaches to manipulation of bubbles and detection of light emitted by them.
Yet, the last three decades has also seen a dramatic progress in the fields of
medicine, chemistry and pharmacology, where, for example, bubbles have been
used as ultrasound contrast agents and as a means of targeted drug delivery
(see Sec.~\ref{sec:1}). Alongside the advances in the adjacent areas of
sonochemistry, sonoprocessing and sonoporation \cite{Mas03, Tan11, Hel16, Yan20},
these factors explain the recent surge in the interest in sonoluminescence from
the point of view of biophotonics, imaging and drug discovery
\cite{Beg19, Can20, Vig20, Boy20}.
\begin{figure}
  \includegraphics[width=.55\textwidth]{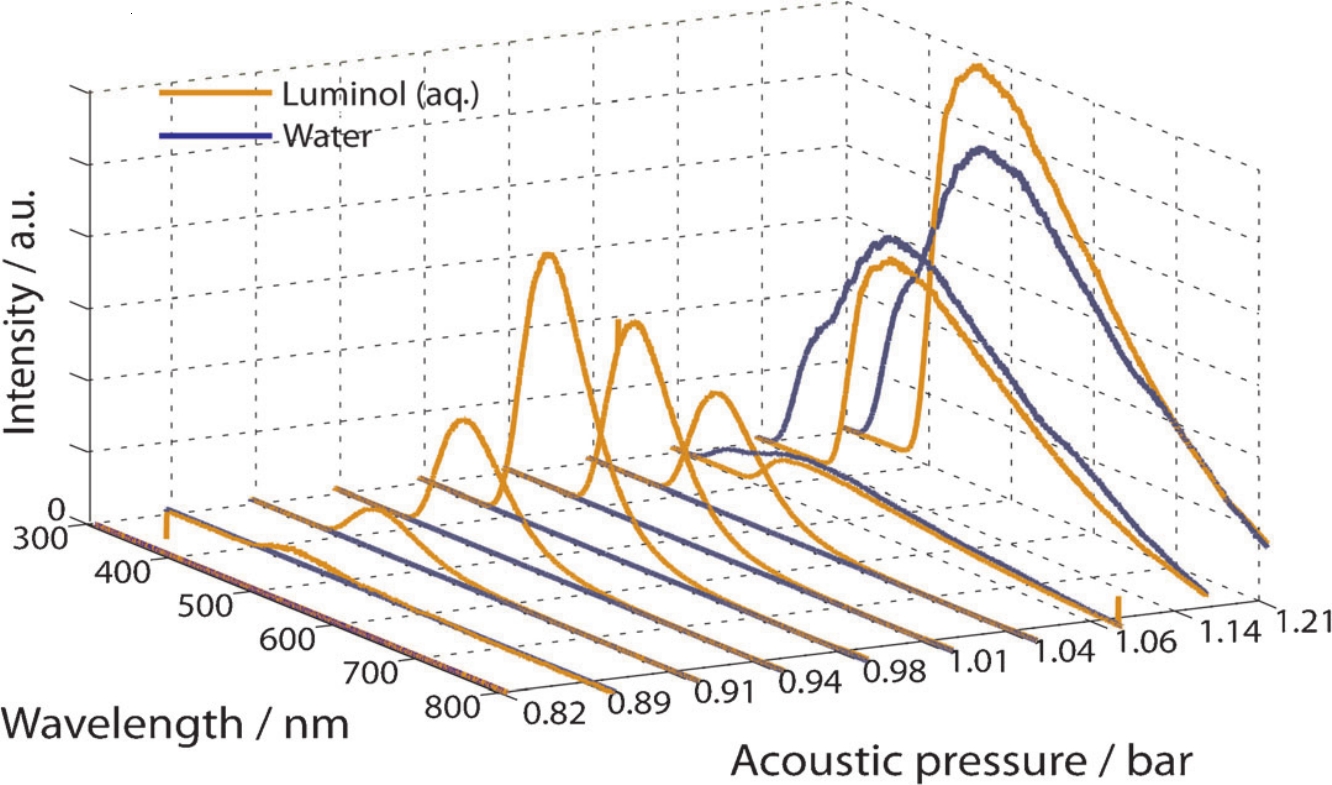}
  \caption{
    Single-bubble sonoluminescence emission spectra obtained in degassed water
    compared with the sonochemiluminescence spectra from an aqueous luminol
    solution. All spectra are plotted as a function of the driving acoustic
    pressure. Reproduced from \cite{Bro12} with permission of John Wiley and
    Sons.~Copyright~2012.\label{Fig_luminol}}
\end{figure}  

We start with the discussion of a sonoluminescence regime that can be characterised 
as an intermediate type of bubble cavitation that bears signatures of both single
and multibubble behaviour. In particular, this regime exploits the property of an 
oscillating bubble to `dance' \cite{Doinikov_book}, i.e.~to undergo erratic spatial
translation movement on a microscopic scale due to Bjerknes force \cite{Bje06, Cru75}
(primary Bjerknes force is caused by an external acoustic pressure field but secondary
Bjerknes force arises between several bubbles that oscillate in the same acoustic
pressure field). While the optical frequency spectrum of the sonoluminescence signal
produced by a dancing bubble was found to contain emission lines that are also present
in a typical multibubble sonoluminescence spectrum, it was also shown that sonochemical
reactions were possible when luminol---a chemical that fluoresces upon oxidation---was
added to the liquid \cite{Hat02}. However, such reactions were not observed in the case
of single bubble sonoluminescence.

The emission of light from luminol in a single-bubble experiment was also investigated
in \cite{Bro12} at different peak amplitudes of the driving acoustic pressure waves.
The results were compared with the data obtained using bubbles in degassed water
(Fig.~\ref{Fig_luminol}). It was shown that while the behaviour of a single bubble
in water followed a typically observed light emission pattern, where the onset of
strong broadband sonoluminescence occurs at the peak acoustic pressure of about 1.05\,bar
and then the light intensity increases sharply, in a luminol solution a typical
emission band centred at the optical wavelength of 425\,nm was observed at significantly
lower acoustic pressures from 0.90 to 1.05\,bar. It is noteworthy that this form of
sonoluminescence is chemical in nature and hence, generally speaking, distinct from
conventional sonoluminescence since it occurs in the solution surrounding the collapsing
bubble through an oxidative radical process \cite{McM99}. However, this result is important
in the context of the mainstream discussion in this article since it demonstrates that
sonoluminescence-like processes can be amplified using luminescent chemicals and that
bright light emission can be achieved using ultrasound waves of a lower amplitude
compared with sonoluminescence in degassed water. This also means that sonoluminescence
could be achieved under medically relevant ultrasound exposure conditions, where the
acoustic pressure levels are lower that those used in studies of fundamental
sonoluminescence properties.

\begin{figure}
  \includegraphics[width=.5\textwidth]{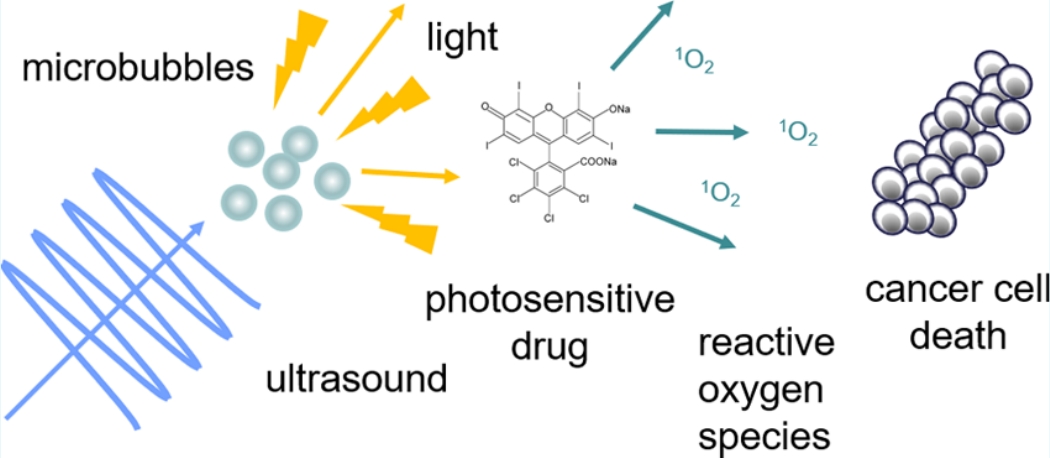}
  \caption{
    Sketch of the operation of a sonoluminescence-activated photosensitive 
    drug. Note that the use ultrasound waves enables employing this approach
    both for diagnostic and therapeutic applications, thereby increasing the range
    of applications of photosensitive drugs that is currently restricted by
    a limited penetration of light in the biological tissues and bodily fluids.       
    Reproduced from \cite{Beg19} under the terms of of a Creative Commons license.
    \label{Fig_Beg19}}
\end{figure}

The ability to achieve multibubble sonoluminescence using therapeutic ultrasound of
a relatively low intensity was also demonstrated in \cite{Beg19}, where it was also
shown that the intensity of so-generated sonoluminescence light could be high enough
to activate a photodynamic therapy (PDT) agent (Fig.~\ref{Fig_Beg19}). The latter
results is particularly important for the field of photodynamic therapy, where it has
been suggested that sonoluminescence and the reactive oxygen species (ROS) associated
with bubble collapse could be employed to activate drugs developed in the framework
of sonodynamic therapy (SDT) \cite{Nim14, Osa16}. STD has demonstrated promising
results for the treatment of aggressive and resistant tumour cell, and it has several
advantages over PDT. For example, while PDT is clinically approved for the treatment
of superficial lesions that can be reached with an endoscope \cite{review_CNBP}, a
large penetration depth of ultrasound into biological tissues, compared with that of
light, enables using SDT in hard-to-reach places in a living body.   

However, it is noteworthy that in the experiment in \cite{Beg19} the authors had to 
use photomultiplier tubes (PMTs) to measure the intensity of sonoluminescence light.
PMTs are a class of vacuum tubes that are extremely sensitive detectors of light due
to their high gain, low noise, ultra-fast response and large area of collection 
\cite{Wri_book}. While PMTs are more complex and expensive compared with
semiconductor-based photodetectors, they remain essential for applications requiring
high-sensitivity detection of imperfectly collimated light, which is the case of
many multibubble sonoluminescence experiments, especially when the amplitude of 
ultrasound waves is not high. Although the use of PMTs is not a limiting factor in
fundamental research studies, it practice it is desirable to increase the intensity
of sonoluminescence light to a level detectable with semiconductor devices. 

Since the amplitude of ultrasound waves used for biomedical applications
cannot be increased beyond certain limits dictated by particular applications
\cite{Mil12}, it has been suggested that low-intensity sonoluminescence light
generated using ultrasound waves of a relatively low amplitude can be effectively
amplified using some of the methods of nanophotonics, where nanoparticles with
certain optical properties have been employed both to increase the intensity
of light and to manipulate light, for example, by changing its wavelength
(colour) \cite{Mak19}. This approach has been validated in the work \cite{Vig20},
where functionalised semiconductor ZnO nanoparticles were employed to increase
the intensity of sonoluminescence light, reduce the cavitation threshold and
promote sonochemical reactions. Combined together, these effects of ZnO
nanoparticles enabled decreasing the amplitude of ultrasound waves needed to observe
bright sonoluminescence. It is also noteworthy that ZnO nanocrystals have been used
in various biomedical fluorescence assays \cite{Hah14}, which additionally justifies
their application in the experiments in \cite{Vig20} and their suitability for 
applications in the fields of biology and medicine.  

ZnO nanoparticles were synthesised using a microwave-assisted synthesis 
technique and then they were amine-functionalised to obtain ZnO-NH$_2$ 
nanoparticles that exhibited stable colloidal suspensions in both ethanol
and distilled water. In the first series of experiments, the sonoluminescence
light emission was detected using a PMT under dark conditions. The cavitation
threshold was recorded by increasing the amplitude of the ultrasound wave 
until the appearance of spikes in the PMT-recorded optical spectrum caused
by sonoluminescence. The so-recorded cavitation threshold in pure water was
about 20\% of the maximum power of the ultrasonic transducer used in the 
measurements and, in agreement with the previous work \cite{Anc20}, it was 
observed that the presence of ZnO-NH$_2$ nanoparticles in the liquid resulted
in a noticeable decrease in the cavitation threshold. 

As a next step, sonoluminescence emission in the presence of ZnO-NH$_2$
nanoparticles was measured using a standard spectrometer capable of detecting 
light in the spectral range from UV to infrared wavelengths. Since the sensitivity
of the spectrometer is lower than that of the PMT device used at the first 
stage of the experimentation, the acoustic pressure was increased fourfold 
(however, it constituted about 80\% of the maximum power deliverable by the 
ultrasound transducer used in the measurements).

\begin{figure}
  \includegraphics[width=.7\textwidth]{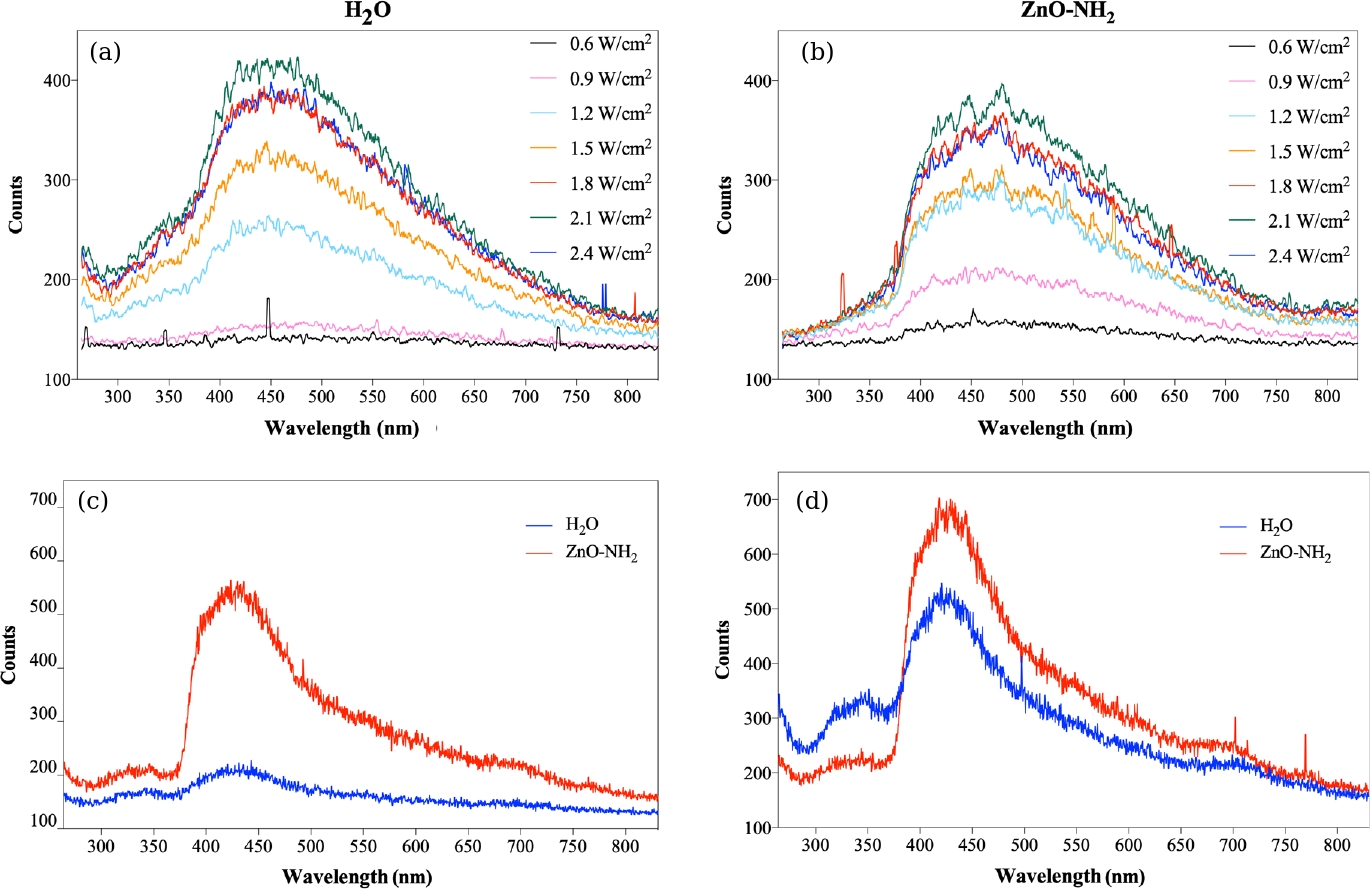}
  \caption{
    Sonoluminescence light spectra in (\textbf{a})~pure water and (\textbf{b})~in
    water containing ZnO-NH$_2$ nanoparticles for different levels of ultrasound 
    intensity.
    (\textbf{c,~d})~The same sonoluminescence light spectra as in panels~(a) and (b)
    but measured in an argon-saturated environment at ultrasound intensities of
    0.9\,W/cm$^2$ and 1.2\,W/cm$^2$, respectively. Note a clear difference between 
    the spectra in the presence and absence of ZnO-NH$_2$ nanoparticles.      
    Reproduced from \cite{Vig20} under the terms of of a Creative Commons license.
    \label{Fig_ZnO}}
\end{figure}

Figure~\ref{Fig_ZnO}(a) shows the sonoluminescence spectra obtained in pure
water as a function of the intensity of the ultrasound wave. One can see a
broad frequency peak that is centred around the wavelength of 450\,nm and covers
the UV and visible spectral ranges. The lowest level of ultrasound intensity 
that leads to the observation of a clear spectrum---1.2\,W/cm$^2$---corresponds
to 40\% of the maximum ultrasound intensity attainable in the experiment 
in \cite{Vig20}. However, in the presence of ZnO-NH$_2$ nanoparticles a clear
sonoluminescence spectrum can be observed at a lower ultrasound intensity of
0.9\,W/cm$^2$ [Fig.~\ref{Fig_ZnO}(b)]. Moreover, one can see that the UV
component of the sonoluminescence light (the wavelength from 250 to 350\,nm)
is absorbed by ZnO-NH$_2$ nanoparticles. Subsequently, it was verified whether
the presence of ZnO-NH$_2$ could result in a complete absorption of light in
the UV range with a concomitant re-emission of light in the visible light region.
To verify this hypothesis, the measurements were repeated in an argon-saturated
atmosphere (it is well-known that the presence of argon results in brighter 
sonoluminescence \cite{Put00, Bre02}). The obtained spectra are shown in 
Fig.~\ref{Fig_ZnO}(c, d) convincingly confirm a strong absorption of UV
light absorption by ZnO-NH$_2$ nanoparticles. However, those results do not 
confirm any significant re-emission since the spectrum of sonoluminescence light 
contains high-amplitude frequency components in the entire visible spectral range
that may obscure the re-emitted light signal. 

Interestingly, it has been shown that multibubble sonoluminescence processes in 
alcohol can be used to synthesise ZnO and ZnO-coated titanium dioxide nanoparticles
\cite{Byu08}. In particular, multibubble sonoluminescence facilitates the supercritical
state of liquid layer, thus promoting high-energy chemical reaction in the layer around
the bubble. Hence, at optimised multibubble sonoluminescence condition in alcohol solutions
containing zinc acetate dihydrate, sodium hydroxide and TiO$_2$, ZnO nanoparticles with
an average diameter of 7\,nm were synthesised \cite{Byu08}.  
 
Carbon nanodots---nanoscale particles of carbon that have been used as a
fluorescent materials that exhibit high biocompatibility and photostability as 
well as low toxicity and eco-friendliness \cite{Reineck, Gor17, Rig20}---are another
type of nanoparticles that can be used to modify the spectrum of sonoluminescence
light \cite{Son21}. In the cited paper, sonoluminescence was achieved in a
single-bubble experimental arrangement, where the emission of light is accompanied
by the sonochemical reaction $H{_2}O \rightarrow H^\bullet + OH^\bullet$. The spectrum
of light emitted by a single air bubble in water typically extends from from UV to
visible wavelengths but its intensity decreases with increasing wavelength, thus
explaining observations of light-emitting bubbles as bright blue spots in otherwise
dark water. Yet, at relatively low levels of driving acoustic pressure waves, a
characteristic optical spectral line corresponding to $OH^\bullet$ appears in the
sonoluminescence spectrum when deionised water is mixed with a noble gas such He,
Ar and Xe \cite{Put00, Bre02}. It is also well-known that certain chemical reactions
between $OH^\bullet$ and other chemicals (e.g.~luminol as demonstrated in the
multibubble experiment in \cite{Riv12}) can help achieve control over sonoluminescence
spectra. In the case of carbon nanodots, in one of the experiments reported in
\cite{Son21} it was demonstrated that while the carbon cores of carbon nanodots are
changed due to a combined mechanical and thermal effects of sonoluminescence, the
chemical groups attached to the surface of carbon nanodots also become modified,
which, in turn, changes the optical properties of the nanodots and shifts the broad
peak of the sonoluminescence spectrum towards the visible range.  

To conclude this section, we review the results of a series of experimental and
theoretical works, where it has been suggested that biocompatible micro- and
nanoparticles made of non-toxic room-temperature liquid metal alloys can be used
to enhance the intensity of the UV component of sonoluminescence light by means
of plasmonic effects \cite{Lu15, Dic17, Dae18, Kal17, Rei19, Boy20}. Surface
plasmons are optical wave that can be resonantly excited at the interface between
a metal and a dielectric material \cite{Raether}. Localised surface plasmon modes
are typically observed at the surface of metal nanoparticles, where the plasmon
resonance frequency depends on the nanoparticle size, its constituent material
and the optical properties of the surrounding environment, but the intensity of
light confined to nanoparticle's surface can be higher than that of the incident
light due to local optical field enhancement processes \cite{Eno12, Mak16, Mak19}.

Nanoparticles suitable for applications in plasmonics are often made of gold and
silver since these two metal exhibit the lowest possible optical absorption losses
compared with other metals such as nickel and iron \cite{Che11, Bon11}. However,
metals other than gold and silver remain valuable in the field of plasmonics because 
their use may help extend the functionality of nanoparticles. Indeed, plasmon resonances
of typical gold and silver nanoparticles lie in the visible and infrared spectral
ranges but their resonance frequencies cannot be readily changed because achieving
this would require modifying the shape, size and constituent material of the 
nanoparticle \cite{Mak16}. Yet, even though there exist techniques for tuning 
the resonance frequency of a nanoparticle using, for example, magnetic fields
\cite{Mak16}, the achievable tuning spectral range is insufficient to shift
the resonance from the visible to UV spectral range. 

To resolve this problem, nanoparticles made of gallium and other metals have been
investigated for their UV plasmonic properties (see \cite{Rei19} and references
therein). In particular, nanoparticles made of gallium alloys stand out for their
intriguing properties that include strong plasmon resonances in the UV spectral range,
non-toxicity and biocompatibility as well as as the fact that certain gallium alloys
can be liquid at room temperature \cite{Lu15, Dic17, Dae18, Kal17}. Subsequently, 
compared with solid-metal nanoparticles that have a constant plasmon resonance 
frequency, the plasmon resonance frequency of liquid gallium alloy nanoparticles can
be readily changed using external electric signals \cite{Dic17}. Yet, it has been
suggested that liquid metal nanoparticles and larger structures such as macroscopic
liquid metal drops could be deformed using acoustic pressure waves \cite{Mak17drop}
and collapsing bubbles \cite{Mak19}. Indeed, let us recall that non-spherically
collapsing bubbles (see the bubble in the middle of Fig.~\ref{Fig_intro}(b) for
illustration) can produce a powerful water jet \cite{Ohl15}. The direction of the
water jet depends on mechanical properties of the surface near which the bubble
collapses. The bubble develops a jet directed towards a rigid surface but away
from a liquid-air interface. Yet, the jet splits near an elastic wall \cite{Fon06, Cur13}.

\begin{figure}
  \includegraphics[width=.7\textwidth]{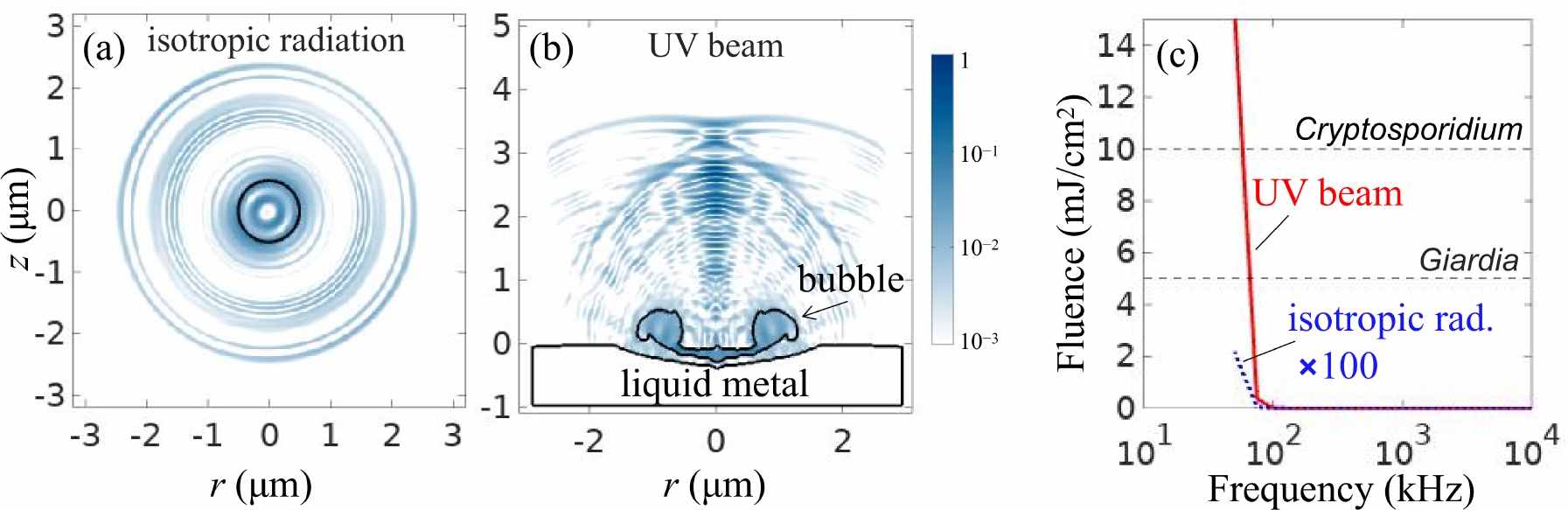}
  \caption{
    Spatial patterns of light emitted by (\textbf{a})~spherical bubble in a bulk 
    of water and (\textbf{b})~non-spherical bubble near the liquid-metal surface.
    The black curves outline the contours of the bubble and liquid-metal surface.
    The equilibrium radius of the bubble is 1\,$\mu$m.
    (\textbf{c})~UV radiation fluence as a function of the ultrasound wave frequency
    $f$. The fluence of isotropic radiation is multiplied by 100. The fluence
    required for the inactivation of {\it Giardia} and {\it Cryptosporidium} pathogens
    is indicated. Reproduced from \cite{Boy20} under the terms of of a Creative Commons
    license.
    \label{Fig_Boyd20}}
\end{figure}

In the work \cite{Boy20}, it was suggested that the collapse of a microbubble near
the surface of a macroscopic liquid-metal drop can result in deformation of the
surface and formation of a concave structure that can serve as a focusing mirror
for UV light. Moreover, it was hypothesises that the bubble collapse can lead to
the emission of sonoluminescence light and that the interaction of the so-generated
light with the liquid-metal mirror can result in the formation of a UV light beam,
where the intensity of the UV radiation is increased manifold due to the plasmonic
properties of the metal when compared with the intensity of light produced as a
result of a spherical bubble collapse. The plausibility of this idea was supported
by a rigorous numerical model of microbubble collapse near a liquid metal surface
\cite{Boy18, Boy19} and a model of sonoluminescence \cite{Bre02} (compare
the result in panels~(a) and (b) in Fig.~\ref{Fig_Boyd20}).

Although potential applications of this plasmonic enhancement effect in the area
of UV photoactivated drugs were not considered in \cite{Boy20}, it has been suggested
that sonoluminescence-based UV beams could be used to irreversibly inactivate most
common pathogens in water. Applying the Fourier transform to spatial patterns of
light created by the collapsed spherical [Fig.~\ref{Fig_Boyd20}(a)] and non-spherical 
[Fig.~\ref{Fig_Boyd20}(b)] bubbles, it was shown that the time-domain optical wave
packets produced by the bubbles carry the spectral radiance for the UV-C wavelengths
$\lambda = 200-280$\,nm. It is well-known that UV-C light can be employed to
inactivate bacteria, viruses and protozoa in environmentally-friendly, chemical-free
and highly effective methods of disinfecting and safeguarding water against
pathogens responsible for cholera, polio, typhoid, hepatitis and other bacterial,
viral and parasitic diseases \cite{Bolton}. As shown in Figs.~\ref{Fig_Boyd20}(a,~b),
a spherical bubble produces an isotropic radiation while its non-spherical counterpart
produces a directed beam perpendicular to the liquid metal surface. Figure~\ref{Fig_Boyd20}(c)
shows that the UV-C light fluence generated by both spherical and aspherical bubbles
dramatically increases at driving ultrasound wave frequencies of less than 100\,kHz.
However, only the formation of a directed beam near the metal microparticles results
in the irreversible inactivation of pathogens. 
  
\section{Sonoluminescence in microfluidics-based systems}
The results reviewed thus far have been achieved in laboratory conditions that are, in
general, different from technologies and equipment employed in modern biomedical
sensing and imaging devices based on microfluidic systems and adjacent approaches.
Microfluidic technologies enable a quick and efficient processing of small quantities of
fluids contained in microscopic channels \cite{Con19}. The reduced turnaround time and
increased productivity with a small device footprint facilitate the application of microfluidic
devices in a variety of experimental situations but their high portability makes them a
technology of choice for point-of-care devices \cite{Has12, Kha12}.

The physical and chemical properties of oscillating and collapsing gas bubbles discussed in
this article have also been essential for the development of microfluidic technologies
\cite{Che09, Rab11, Kha12, Zho13, Ohl15}. Therefore, it is plausible that sonoluminescence
can be achieved in microfluidics-based system. Indeed, in the work \cite{Riv12} micromachined
pits on a solid substrate were employed to nucleate and stabilise microbubbles in a liquid
irradiated with an ultrasound pressure wave, and ultrasound-driven collapse of such bubbles
resulted in the emission of light. In turn, the generated light interacted with luminol
that was present in the liquid, thus producing a sonochemiluminescence effect. The intensities
of sonoluminescence and sonochemiluminescence optical signals were recorded at several 
operating regimes related to the peak acoustic pressure amplitude at the ultrasonic frequency
of 200\,kHz for a pure water environment and for aqueous luminol and propanol solutions.
Several arrangements of pits on the substrate were investigated. At high acoustic pressures
and in the presence of more than one pit on the substrate, the microbubbles combined into a
chain or formed a triangular structure known as a streamer \cite{Ohl15}. In all experiments
the intensity of sonochemiluminescence light was amplified as the number of pits was increased.
This result was reproduceable at both low and high acoustic pressure levels.

\begin{figure}
  \includegraphics[width=.7\textwidth]{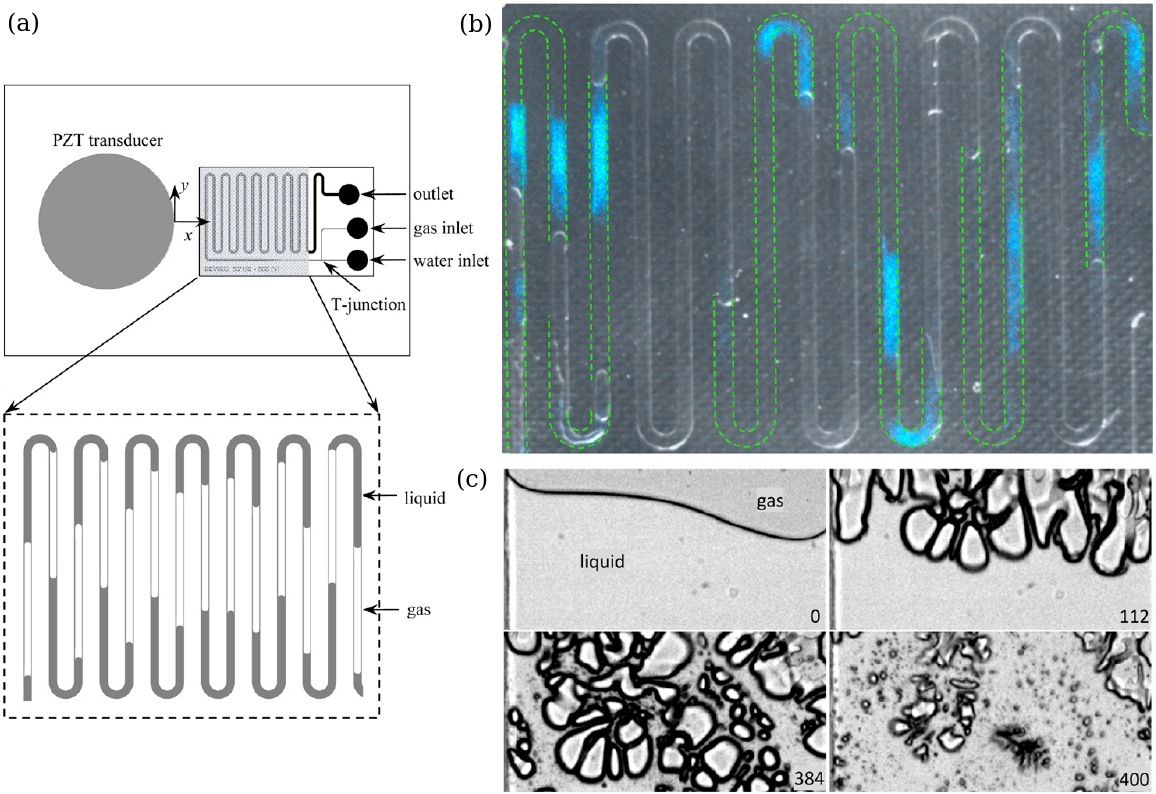}
  \caption{
    (\textbf{a})~Sketch of the microfluidic device designed in \cite{Tan11} to observe
    sonochemistry and sonoluminescence effects. The piezoelectric transducer (PZT) used for
    the creation of the driving ultrasound wave is glued to a glass substrate to which a
    meandering microfluidic channel is bonded. The inlets consist of two ports for gas and
    water, respectively. The inset shows a typical distribution of gas and liquids along
    the microchannel.
    (\textbf{b})~Picture of the microfluidic channel geometry superimposed on a photograph 
    showing a sonochemiluminescence activity due to the oxidation of luminol. The green dashed
    lines indicate the positions of liquid slugs inside the channels just before the ultrasound
    was switched on.
    (\textbf{c})~Sequence of images taken with a high-speed camera showing the bubble distribution
    near a gas-liquid interface. The number in each frame is the time in microseconds after 
    switching on the ultrasound signal. One can see that the initial gas-liquid interface leads
    to surface instabilities, expansions of large bubbles and eventually to bubble collapses.
    The interface was exposed to a harmonic ultrasound driving wave with the frequency of 103.6\,kHz.
    The width of each frame is 500\,$\mu$m, which is also the width of the microfluidic channel.
    Reproduced from \cite{Tan11} under the exclusive PNAS License to Publish. Copyright~2011
    National Academy of Sciences.\label{Fig_microfluid}}
\end{figure}

In another relevant work \cite{Tan11}, the conditions suitable for a realisation of
sonoluminescence and sonochemistry were created using bubbles confined within a narrow
channel of a microfluidic device [Fig.~\ref{Fig_microfluid}(a)], where the bubbles assume a
planar pancake-like shape. As a result of collapse of such bubbles the authors observed the
formation of $OH$ radicals and the emission of sonoluminescence light [Fig.~\ref{Fig_microfluid}(b)].
The observation of chemical reactions was closely linked to the gas-liquid interfaces
[Fig.~\ref{Fig_microfluid}(c)] and it was suggested that spatial control over them could 
be used to control sonochemical reactions. Furthermore, the decay time of the light emitted
from the sonochemical reaction was several orders faster than that in the bulk of liquid
and the emission of light vanished immediately as the ultrasound was switched off (also 
see a relevant discussion of the Purcell effect below). 

It has also been shown theoretically that microscopic water-core optical fibre can be used 
as microfluidic channels, where bubbles may oscillate, collapse and emit light under the 
pressure of an ultrasound pressure wave \cite{Mak17}. The fibre analysed in \cite{Mak17}
consisted of a Teflon tube filled with water. The optical refractive index of Teflon is
slightly lower than that of water, which creates the conditions suitable for efficient 
concentration of light inside the water core. Although optical losses in the water-core
fibre are higher than those in modern solid-state fibres, the liquid-based system has certain
advantages relevant to the main discussion in this article. In particular, it was demonstrated
that the water-core fibre can efficiently guide light when the Teflon tube is immersed into
another liquid. Yet, numerical simulations conducted in \cite{Mak17} revealed that ultrasound
waves can pass through the structure of the water-immersed fibre with negligibly small
acoustic scattering and attenuation. Given these unique properties, it is plausible that
sonoluminescence light produced inside the core of the fibre could be efficiently collected
and detected using a standard photodetector connected to the output end of the fibre.    

\section{Conclusions and open questions}
In this article, we have reviewed the recent results in the emergent field of fluorescence
and plasmonic nanoparticle-enhanced sonoluminescence with a focus on potential applications
of light emitted by collapsing bubbles in the fields of photonics and biophotonics and 
adjacent areas. Given the ability of ultrasound waves that cause the bubble collapse to 
readily pass through the skin and bodily fluids and tissues, and due to the presence of natural
and artificial gas bubbles inside a living body, it has been argued that sonoluminescence light
could be used to activate certain drugs inside the body, also providing a means for inactivation
of pathogens with UV radiation and for imaging of individual cells without application of
chemicals that can pose ethical issues. While rigorous {\it in vivo} tests of these potential
applications of sonoluminescence have not been conducted yet, the results obtained in a number
of recent experimental works, where commercial medical ultrasound imaging microbubble contrast
agents and medical-grade ultrasound wave intensities were employed, strongly speak in favour of
plausibility of the concept of sonoluminescence-based photonics and biophotonics.

However, extra research studies are required for sonoluminescence-based processes to attract the
interest of a large number of scientists working on photonics and biophotonics. Here, we outline
the current challenges and open questions the resolution of which, in our opinion, will shape the
future research efforts in the area of sonoluminescence and gas bubble photonics.

Firstly, the experiments involving the interaction of sonoluminescence light with nanoparticles
have thus far been conducted in purified liquids and liquids that contain chemicals that 
facilitate observations of bright sonoluminescence. However, in biologically-relevant situations
the composition of fluids may be unfavourable for the production of sonoluminescence from the
point of view of both fluid dynamics and optics. Therefore, additional studies of sonoluminescence
have to be conducted in different types of biofluids or fluids that have similar properties, and
the research questions such as the impact of the biofluid viscosity and of light absorption and
scattering have to be rigorously addressed. Indeed, while in single-bubble sonoluminescence
experiments microbubbles can be trapped in water using standing acoustic pressure waves,
achieving a stable bubble position in liquids with higher viscosity has proven to be challenging
since in such liquids bubbles undergo a quasiperiodic circular translational motion on the time
scale of seconds \cite{Toe06}.

Secondly, it is known that the intensity of light emitted by a single sonoluminescing bubble
can be increased using several driving ultrasound waves of different frequencies. The
optimal values of pressure amplitude and relative phase of the ultrasound waves were reported
in \cite{Lu03}, where a two harmonic frequency bubble forcing regime ---26.5\,kHz and
53\,kHz---has been shown to result in increased temperatures inside a sonoluminescing bubble 
that maintained its spherical stability despite the action of two acoustic pressure waves.
Subsequently, a combination of this sonoluminescence regime with nanoparticles should allow
achieving higher sonoluminescence light intensities. It is also plausible that the use of
more than two excitation frequencies or even a quasi-continuum of frequencies would result 
in strong sonoluminescence provided that the bubble maintains its spherical shape during the 
oscillation. One potential test of this idea could employ acoustic frequency combs---signals 
with spectra consisting of equidistant frequency peaks \cite{Mak22}. Despite a multifrequency
content of an acoustic frequency comb signal, a bubble driven by it can still maintain its
spherical shape \cite{Mak21} that is favourable for producing sonoluminescence \cite{Bre02}.     

Thirdly, in the field of nanophotonics it is well-established that luminescent quantum emitters
have relatively long emission lifetimes of several tens of nanoseconds and non-directional emission
patterns. These intrinsic optical properties are not suitable for applications in many nanophotonic
devices such as single-photon sources \cite{Mak10}, where fast radiative rates are required for operation
at high frequencies but directionality is needed to achieve a high efficiency of light collection
\cite{Lalanne, Aks14}. One of the approaches to the resolution of this problems exploits the Purcell
effect---a modification of the spontaneous emission rate of an emitter of light in the presence of
a resonant cavity \cite{Lalanne, Aks14, Kra15}. The Purcell effect has also been observed in systems
consisting of plasmonic nanoparticles---nanoantennas \cite{Lalanne, Aks14, Kra15, Mak16}---that can
modify the photonic environment of fluorescent sources of light by enhancing the spontaneous emission
rate.

Given that the physics of sonoluminescence can also be explained in terms of quantum radiation
\cite{Ebe96, Ebe96_1}, it is conceivable that the properties of sonoluminescence-based light 
sources could also be controlled using the Purcell effect originating from the plasmonic properties
of metal nanoparticles. Moreover, such tests could be conducted in opto-microfluidic channels used
in biomedical and pharmaceutical research, where, for example, sonoluminescence of bubbles nucleated
in pits formed on a silicon substrate has been reported in \cite{Riv12}. Due to the decent optical
properties of silicon, the pits formed in this material posses the properties of optical micro-cavities,
where a strong Purcell effect has been observed \cite{Lalanne, Aks14, Kra15}. An enhancement of the
sonoluminescence light intensity could also be investigated in an acousto-optical device \cite{Mak_extr},
where a microscopically patterned metallic film simultaneously plays the roles of a bubble-generating
substrate and of an optical resonator \cite{Mak10}. Furthermore, the Purcell effect could be used to 
inhibit the emission of a source of light \cite{Lalanne}. Subsequently, from the point of view of 
the fundamental physics, one might be interested in studying potential effects of light emission 
inhibition under otherwise favourable for the production of sonoluminescence conditions. Yet, since
the existence of acoustic and elastic counterparts of the optical Purcell has been demonstrated
\cite{Lan18, Sch18}, and the possibility of combining the optical and acoustic Purcell effects has 
been pointed out \cite{Sch_wombat}, one might control sonochemiluminescence light intensity using 
both optical and acoustic techniques.     

It is also worth noting that sonoluminescence processes may be of interest to the burgeoning field
of metamaterials \cite{Zhe12, Pod13}. In general, a metamaterial is any artificial material engineered
to have a property that is not found in naturally occurring media. For instance, an electromagnetic
metamaterial affects electromagnetic waves that interact with its structural features. Since a typical
size of the features of a metamaterial is smaller than the wavelength of the incident wave, photonic
metamaterials that operate at optical frequencies are usually structured at the nanometre scale \cite{Zhe12}.
Similarly, the features of acoustic metamaterials designed to control, direct and manipulate acoustic waves
in fluids and solids are typically smaller than the wavelength of the incident acoustic wave \cite{Tri02}.

Among different kinds of metamaterials, hyperbolic metamaterials have attracted a special attention 
since their hyperbolic dispersion relationship gives rise to useful effects including enhancement of
spontaneous emission, negative refraction and ability to overcome the optical diffraction limit \cite{Pod13}.
The concept of hyperbolic metamaterials has been extended to waves other than electromagnetic waves
and light, where, in particular, it has been suggested that acoustic waves existing in hyperbolic
metamaterials can be regarded as analogues of gravitational waves \cite{Smo22}. In this context and
in light of the theoretical works \cite{Ebe96, Ebe96_1} demonstrating a quantum nature of sonoluminescence,
it has been suggested that the efficiency of sonoluminescence could be greatly enhanced using hyperbolic
metamaterials and that studies of sonoluminescence processes in hyperbolic metamaterials could help
search for quantum gravity effects \cite{Smo22}. One suitable candidate for such experimental tests 
is a ferrofluid-based self-assembled hyperbolic metamaterial, where a chain of cobalt nanoparticles,
which leads to formation of the hyperbolic metamaterial structure, and microbubbles are formed inside
the ferrofluid after application of external magnetic field \cite{Smo17}.

\vspace{6pt} 

\funding{This work has been supported by the Australian Research Council
through the Future Fellowship (FT180100343) program.} 

\acknowledgments{ISM thanks Professor Sergey Suslov, Dr. Andrey Pototsky and
Mr. Bui Quoc Huy Nguyen for their support and invaluable discussions.}

\conflictsofinterest{The author declares no conflict of~interest.} 

\abbreviations{Abbreviations}{
The following abbreviations are used in this manuscript:\\

\noindent 
\begin{tabular}{@{}ll}
    blood-brain barrier &BBB\\
    {\it in~vitro} fertilisation &IVF\\	
    photomultiplier tube &PMT\\
    photodynamic therapy &PTD\\
    piezoelectric transducer &PZT\\
    reactive oxygen species &ROS\\
    sonodynamic therapy &SDT\\
    ultraviolet &UV\\
    
\end{tabular}
}

\end{paracol}
\reftitle{References}


\externalbibliography{yes}
\bibliography{refs}

\end{document}